\documentclass{article}
\usepackage{amsmath,amssymb}
\newcommand{\bra}[1]{\langle #1\vert}
\newcommand{\ket}[1]{\vert #1\rangle}

\newcommand{\eg}{e.\,g.\ }
\newcommand{\ie}{i.\,e.\ }
\newcommand{\iden}{{\bf 1}}
\newcommand{\Tr}{\mbox{\rm Tr}}
\begin{document}

\title{Leggett's theorem without inequalities\footnote{An expanded version of this article will be made available on the PHILSCI archive (http://philsci-archive.pitt.edu/).}}\author{Guido 
Bacciagaluppi\footnote{Centre for Time, Department of Philosophy, University of Sydney, and Institut d'Histoire et de Philosophie des Sciences et des Techniques (CNRS, Paris 1, ENS), Paris. Address for correspondence: Department of Philosophy, University of Sydney, NSW 2006, Australia (e-mail: guidob@univ-paris1.fr).}} 
\date{}
\maketitle

\begin{abstract}
We prove a no-go theorem for a class of hidden variables theories that satisfy parameter independence. Specifically, we show that, assuming two conditions, there are no non-trivial hidden variables models of the quantum predictions for product measurements on two systems in any maximally entangled state in a Hilbert space of dimension at least $3\times 3$. The two conditions are parameter independence and a condition that we call conditional parameter independence. The result is analogous to the recent no-go theorems based on Leggett's inequalities and their generalisations.
\end{abstract}

\section{Introduction}
The main assumption in the standard derivations of the Bell inequalities is Bell's factorisation condition on the joint probabilities for outcomes of an EPR-type experiment (Bell, 1971). The best-known analysis of the meaning of this condition was given by  Jarrett (1984) and by Shimony (1986), and interprets factorisation as the conjunction of two conditions, called outcome independence (OI) and parameter independence (PI) by Shimony (or completeness and locality, respectively, by Jarrett).\footnote{Shimony's and Jarrett's conditions are not identical, but the differences will be largely irrelevant in the following. See the remarks in the next section.} These two conditions in turn capture the idea that:
    \begin{itemize}
      \item[(OI)] given a complete description of the state of the system and the experimental context, the 
    outcomes on the two sides of an EPR experiment should be statistically independent;
      \item[(PI)] given the complete state of the system, the probabilities for the outcomes on each side of    
      an EPR experiment (say, Bob's) should be independent of the experimental context on the other side  
    (say, Alice's).
  \end{itemize}

The general view is that PI is a locality requirement, since its violation would allow signalling across the wings of the experiment if one could fix the values of the hidden variables, while violations of OI (or at least violations of OI alone) are taken to be compatible with special relativity and indicative rather of some form of holism or non-separability, and in this sense allow for a `peaceful coexistence' (Shimony, 1978) between quantum mechanics and special relativity.

According to this view, one might be able to construct non-trivial hidden variables theories that coexist peacefully with special relativity. Recent work, however, has uncovered constraints on the possibility of hidden variables theories that satisfy PI. This work centres around Leggett's (2003) inequalities and their further theoretical elaborations and experimental tests (cf.\ Branciard {\em et al}., 2008).  In this paper, we shall prove a theorem analogous to (and in certain ways generalising) this work, and in the process arguably make more explicit some of the assumptions involved.

Specifically, we shall show that any hidden variables theory that makes non-trivial predictions (\ie improves on the quantum predictions if the hidden variables $\lambda$ are known) for maximally entangled states in dimension $3\times 3$ or higher and that satisfies both PI and a further condition, which we call {\em conditional parameter independence} (CPI), is incompatible with the predictions of quantum mechanics. 

CPI is closely related to the condition of `constrained locality' introduced (for a different purpose) by Jones and Clifton (1993), and reads informally as:
  \begin{itemize}
    \item[(CPI)] given the complete state of the system, the probabilities for the outcomes on each side of 
    an EPR experiment, {\em conditional on the outcomes on the other side}, should be independent of 
    the experimental context on the other side.
  \end{itemize}

Our result shows that, if one holds fast to PI, a hidden variables theory will either have to become trivial in modelling the predictions of quantum mechanics for maximally entagled states, or will have to violate CPI, which in our opinion amounts to a new peculiar form of contextuality.  
  
The paper is structured as follows. We first recall the standard framework of hidden variables theories in section~\ref{framework}. Then we state and prove our main theorem in section~\ref{theorem}. The final section~\ref{Leggett} spells out the relation to Leggett's theorem.

\section{The framework of hidden variables}\label{framework}
We shall presuppose the usual formal framework of hidden variables theories as used in the discussion of the Bell inequalities. In such a theory, the complete state of a system is given at the individual level by a pair
  \begin{equation}
    (\rho, \lambda )\ ,
    \label{eq1}
  \end{equation}
where $\rho$ is the quantum state and $\lambda$ is called a hidden variable. At the statistical level, one will have some distribution $\rho(\lambda)$ over the hidden variables. Typically, $\rho(\lambda)$ will depend on the quantum state $\rho$, but not on other factors such as the choice of the measurements to be performed on the system (which is generally assumed to be specifiable independently at the initial time).

Once the full experimental context is given, the complete state of the system determines the probabilities for the outcomes of measurements of all quantum mechanical observables. For present purposes, we can restrict ourselves to observables represented by projection-valued measures. We shall thus take a hidden variables theory to determine all probabilities of the form:
  \begin{equation}
    p^{i_P}_{(\rho,\lambda)}(P)\ ,
    \label{eq2}
  \end{equation}
for all possible states $(\rho,\lambda)$, projectors $P$ and measurement contexts $i_P$ suitable for measuring $P$. In the case of measurements of product projections $P\otimes Q$, we shall assume that the experimental context can be thought of as separable into local contexts $i_P$ and $i_Q$ pertaining to the two wings of the experiment. 

As to whether one should include the microscopic degrees of freedom of the apparatus, including any `apparatus hidden variables', in $\lambda$ or in the experimental contexts, this ambiguity is of no essential importance in the following. For convenience, we shall adopt the former choice, because in this case we can take $i_P$ to be in principle completely fixed by the experimenter and need not explicitly consider also probability distributions over the contexts.\footnote{In this case $\rho(\lambda)$ could in principle include `conspiratorial' correlations between `system hidden variables' and `apparatus hidden variables'. In order to derive the Bell inequalities, these correlations need to be ruled out separately (or one can average over the apparatus hidden variables).}

Given the above, it is required of a hidden variables theory that averaging over the hidden variables must reproduce the statistical predictions of quantum mechanics:
  \begin{equation}
    p_\rho(P)=\int p^{i_P}_{(\rho,\lambda)}(P)\rho(\lambda)d\lambda\ ,
    \label{eq3}
  \end{equation}
\ie averaging over $\lambda$ with the distribution $\rho(\lambda)$ yields the quantum mechanical probabilities
  \begin{equation}
    p_\rho(P)=\Tr (\rho P)\ ,
    \label{eq4}
  \end{equation}
which are in fact independent of any specific context $i_P$ (besides the fact that it is appropriate to measuring $P$).  

A hidden variables theory, finally, will be said to be trivial (or trivial for some set of states $\rho$) if it makes the same probabilistic predictions as quantum mechanics even for every single $\lambda$, \ie if the complete states $(\rho, \lambda)$ do not improve on the statistical predictions of the quantum states $\rho$.

\section{The theorem}\label{theorem}
Let us state the precise definitions of the conditions mentioned in the introduction. 
{\em Outcome independence} (OI) is the condition that:
  \begin{equation}
    p^{i_P,i_Q}_{(\rho,\lambda)}(P,Q)=p^{i_P,i_Q}_{(\rho,\lambda)}(P)p^{i_P,i_Q}_{(\rho,\lambda)}(Q)\ .
    \label{eq5}
  \end{equation}
{\em Parameter independence} (PI) requires that:
  \begin{equation}
    p^{i_P,i_Q}_{(\rho,\lambda)}(P)=p^{i_P}_{(\rho,\lambda)}(P)\qquad\mbox{independently of $i_Q$}
    \label{eq6}
  \end{equation}
(and analogously for $Q$ independently of $i_P$). 
Finally, {\em conditional parameter independence} (CPI) requires that:
  \begin{equation}
    p^{i_P,i_Q}_{(\rho,\lambda)}(P|Q)=p^{i_P}_{(\rho,\lambda)}(P|Q)\qquad\mbox{independently of $i_Q$}
    \label{eq7}
  \end{equation}
(and analogously for $Q|P$ independently of $i_P$).

We shall now prove the following 

\noindent{\bf Theorem}:
  {\em There is no hidden variables theory satisfying PI and CPI that makes non-trivial predictions for measurements of product  
  projections on two $n$-dimensional systems, with $n$ finite and 
  greater or equal 3, in a maximally entangled state.}  

The theorem follows from four separate results. The first is
 
  \noindent{\bf Proposition 1}:
  {\em Take a hidden variables model of the quantum predictions for a maximally entangled state in 
  dimension $2\times 2$ or higher, and impose PI. It follows that the {\em marginal} probabilities of the 
  model must be {\em non-contextual}.}

Indeed, it is a property of maximally entangled states that for any projection $P$ of Alice's system, there is a projection $Q$ of Bob's system, such that $P$ and $Q$ are perfectly correlated. That is, 
  \begin{equation}
    p_\Psi(P)=p_\Psi(P,Q)=p_\Psi(Q)\ .
  \end{equation}
We shall now fix $\Psi$ and drop the index $\Psi$ (or $\rho$) from the formulas.

It follows that, in order for the hidden variables theory to reproduce this prediction on average, one must have
  \begin{equation}
    p^{i_P,i_Q}_\lambda(P)=p^{i_P,i_Q}_\lambda(P,Q)=p^{i_P,i_Q}_\lambda(Q)\ ,
  \end{equation}
also for every individual $\lambda$, $i_P$ and $i_Q$.

But then, $p^{i_P,i_Q}_\lambda(P)=p^{i_P,i_Q'}_\lambda(P)$ implies
$p^{i_P,i_Q}_\lambda(Q)=p^{i_P,i_Q'}_\lambda(Q)$, i.e. imposing PI enforces independence of the local experimental context. Again because of PI, this holds independently of whether or not $P$ is in fact measured alongside $Q$.

Proposition 1 is merely a variant of the non-locality arguments for contextual hidden variables theories (taken in the sense of 0-1 valuations on the projections in Hilbert space) that have been discussed several times in the past, notably by Stairs (1983), Heywood and Redhead (1983),  Brown and Svetlichny (1990) and Bacciagaluppi (1993). Indeed, taking the special case of all probabilities having values 0 or 1, and of dimension $3\times 3$ or higher, Proposition 1 yields a non-contextual value assignment to all projections on Alice's (or Bob's) side, which is impossible by the Kochen-Specker theorem or the corollary to Gleason's theorem. Therefore, one obtains a proof of non-locality for deterministic hidden variables theories without the use of inequalities.

In our (probabilistic) case, applying the result to dimension $3\times 3$ or higher, one can invoke Gleason's theorem and conclude that the `hidden' marginals must be quantum mechanical in form, i.e. given by density operators in some convex decomposition of the completely mixed state. 

In order to obtain incompatibility with quantum mechanics we therefore have to impose some further condition, namely CPI:
  \begin{equation}
    p^{i_P,i_Q}_{(\rho,\lambda)}(P|Q)=p^{i_P}_{(\rho,\lambda)}(P|Q)\ .
    \label{final1}
  \end{equation}
Evidently, CPI is similar in form to PI, but the two are logically independent. Indeed, it is easy to imagine examples in which CPI is satisfied but PI is violated (this possibility plays an important role in Jones and Clifton (1993)). Conversely, one can easily imagine examples in which CPI is violated, but the dependence on the nearby context is washed out if one averages over the nearby outcomes, so that PI is satisfied. 

Note that, if one assumes OI, then PI and CPI are trivially equivalent. This means in particular that, if one assumes PI, a violation of CPI implies a violation of OI, so that hidden variables theories satisfying PI but violating CPI indeed violate the assumptions of Bell's theorem.
We shall now prove the following
  
  \noindent{\bf Proposition 2}: {\em Take any probability distribution on the product projections in dimension $2\times 2$ or higher having
  non-contextual marginals. Imposing CPI will then force non-contextuality of the {\em joint} distribution.}
  
The proof is trivial, since imposing non-contextuality on the marginal probabilities and on the conditional probabilities obviously enforces that of the joint probabilities. Explicitly, assuming CPI, for any such probability distribution, say, $p_\lambda^{i_P,i_Q}(P,Q)$ we have:
  \begin{equation}
    p_\lambda^{i_P,i_Q}(P,Q)=p_\lambda^{i_P,i_Q}(P|Q)p_\lambda^{i_P,i_Q}(Q)=p_\lambda^{i_P}(P|Q)p_\lambda(Q)\ ,
  \label{eq15}
  \end{equation} 
where the first factor on the right-hand side is independent of $\lambda$ because of CPI, and the second because of the non-contextuality of the marginals. Similarly,  
  \begin{equation}
    p_\lambda^{i_P,i_Q}(P,Q)=p_\lambda^{i_P,i_Q}(Q|P)p_\lambda^{i_P,i_Q}(P)=p_\lambda^{i_Q}(Q|P)p_\lambda(P)\ .
  \label{eq16}
  \end{equation}
By (\ref{eq15}) and (\ref{eq16}), the joint probabilities are independent of both $i_P$ and $i_Q$.

Thus, while in dimension $3\times 3$ or higher PI constrains the marginals $p_\lambda^{i_P,i_Q}(P)$ and $p_\lambda^{i_P,i_Q}(Q)$ to be non-contextual, it does not constrain in the same way the joint probabilities $p_\lambda^{i_P,i_Q}(P,Q)$. The joint probabilities could still depend explicitly on the contexts $i_P$ and $i_Q$, since it is obvious that fixing the marginals of a probability distribution does not fix the distribution itself. Different pairs of local contexts $(i_P,i_Q)$ could determine different joint distributions, all compatible with the same non-contextual marginals. CPI rules out precisely this peculiar form of contextuality. 

We shall see, however, that the only non-contextual probability distribution of this kind that reproduces the perfect correlations of a maximally entangled quantum state is that given by the quantum state itself.

The next step in the proof is a rather well-known generalisation of Gleason's theorem for the case of 
product projections (see \eg Barnum {\em et al}., 2005, esp. section~4, and references therein):

  \noindent{\bf Proposition 3}: {\em Take any probability distribution on the product projections on ${\cal H}\otimes{\cal H}$ in dimension
  $3\times 3$ or higher, say $p_\lambda(P,Q)$. Then there is a self-adjoint operator $\Lambda$ on ${\cal H}\otimes{\cal H}$, with 
  $\Tr(\Lambda)=1$, such that}
    \begin{equation}
      p_\lambda(P,Q)=\Tr(\Lambda P\otimes Q)\ .
      \label{eq16a}
    \end{equation}
  {\em Equivalently, there is an affine and positive map $\Phi_\Lambda$ from ${\cal H}$ to itself, with 
  $\Tr(\Phi_\Lambda(\iden))=1$, such that}
    \begin{equation}
      p_\lambda(P,Q)=\Tr(P\Phi_\Lambda(Q))\ .
      \label{final3}
    \end{equation}
Intuitively, the positive operator $\Phi_\Lambda(Q)$ is the state on 
Alice's side after Bob has performed a selective measurement of $Q$; indeed, it is an (unnormalised) 
density matrix. Note that the operator $\Lambda$ itself need not be positive (\ie the 
expression (\ref{eq16a}) need not be positive other than for product projections). For the special 
case of probability distributions corresponding to quantum states, $\Lambda$ is of course 
a density matrix, and the conclusion holds also in dimension $2\times 2$.  

Proposition 3 establishes that the hidden states $\lambda$ of a model of EPR-type measurements 
satisfying both PI and CPI must be equivalent to such operators $\Lambda$. 
Of course, if one wished to construct theories of this form, one would need to worry how one could extend (contextually) the probability measures of the form $\Tr(\Lambda P\otimes Q)$ to the full set of projections while maintaining positivity. The question is moot, however, because for maximally entangled states there are no such non-trivial $\Lambda$. 
This is the content of
 
  \noindent{\bf Proposition 4}: {\em Let $\rho$ be a maximally entangled state in dimension $2\times 2$ or higher. Then, $\rho$ cannot be 
  decomposed as a convex sum of the form}
    \begin{equation}
      \rho=\eta\Lambda_1+(1-\eta)\Lambda_2\ ,
      \label{final4}
    \end{equation}
  {\em with $0<\eta<1$ and $\Lambda_1$ and $\Lambda_2$ two self-adjoint operators as in Proposition~3.}

For dimension $2\times 2$, given that $\rho$ is pure, the claim follows from an old result by Choi, 
namely that the extremal points in the set of all $\Lambda$ are the pure quantum states and their 
partial transposes (see Barnum {\em et al.}, 2005). The proof of Proposition~4 for the general case will be postponed to 
the appendix.\footnote{The original proof of this conjecture was suggested to me by Howard Barnum 
and Alex Wilce (private discussion, College Park, 26 April 2008), to both of whom I am extremely 
grateful. The proof given in the appendix uses many of the same ingredients (in particular the crucial 
Lemma~2), but it is more elementary.}

Putting together Propositions~1--4, we obtain our desired result. Indeed, take a maximally entangled state $\rho$ in dimension $3\times 3$ or higher, and consider possible hidden variables models for the corresponding quantum probabilities. From Proposition~1, assuming PI, we show that the marginals of the hidden distributions are non-contextual. From Proposition~2, assuming CPI, we show that the joint hidden distributions are non-contextual. From Proposition~3, the hidden states are equivalent to operators of the form $\Lambda$, and from Proposition~4, $\Lambda=\rho$ for all hidden states. That is, there is no non-trivial hidden variables model satisfying PI and CPI for maximally entangled states in dimension $3\times 3$ or higher.

Note that in dimension $2\times 2$ the theorem fails. A counterexample is provided by the model of the singlet state probabilities constructed using Popescu-Rohrlich non-local boxes (Cerf {\em et al.}, 2005). The possibility of such a model, however, is limited to dimension $2\times 2$.

\section{Leggett's theorem without inequalities}\label{Leggett}
Leggett (2003) has shown that a certain class of hidden variables theories in dimension $2\times 2$ that violate the assumptions of Bell's theorem can be nevertheless ruled out through another inequality, which is satisfied by these theories but violated by quantum mechanics.  
In what follows we shall base ourselves on the modification and generalisation of Leggett's results given by Branciard {\em et al.} (2008). 

The relevant class of hidden variables theories is defined as follows. Each $\lambda$ is given by a pair of vector states $\mu,\nu$, and the corresponding probability distributions for the outcomes of product measurements have the form
  \begin{equation}
    p^{{\bf a},{\bf b}}_{\mu\nu}(\alpha,\beta)=\frac{1}{4}\Big(1+\alpha\langle{\bf a}{\bf\sigma}\rangle_\mu
                                                                            +\beta\langle{\bf b}{\bf\sigma}\rangle_\nu
                                                                            +\alpha\beta C_{\mu\nu}({\bf a},{\bf b})\Big)\ .
     \label{final10}                                                                       
  \end{equation}
Such distributions return the quantum mechanical marginals for $\mu$ and $\nu$, but allow for arbitrary correlation coefficients $C_{\mu\nu}({\bf a},{\bf b})$. They have no context dependence whatever, so satisfy PI (and in fact CPI), but allow for violations of OI.\footnote{In this notation, $\alpha$ and $\beta$ label the results for measurements in the given (signed) directions ${\bf a},{\bf b}$. Since these results flip signs if the directions do, there is no dependence on the sign of ${\bf a}$ and ${\bf b}$ if the outcomes are labelled by Hilbert-space projections (as done in the rest of this paper).} 

Imposing positivity on (\ref{final10}) leads to inequalities of the type
  \begin{equation}
    \frac{1}{3}\sum_{i=1}^3|C({\bf a}_i,{\bf b}_i)+C({\bf a}_i,{\bf b}'_i)|\leq2-\frac{2}{3}|\sin\frac{\varphi}{2}|\ ,
    \label{final11}
  \end{equation}
for judiciously chosen directions ${\bf a}_i,{\bf b}_i,{\bf b}'_i$, and where $\varphi$ is the angle between the directions ${\bf b}_i$ and ${\bf b}'_i$. The corresponding quantum prediction is
  \begin{equation}
    \frac{1}{3}\sum_{i=1}^3|C({\bf a}_i,{\bf b}_i)+C({\bf a}_i,{\bf b}'_i)|=2|\cos\frac{\varphi}{2}|\ ,
    \label{final12}
  \end{equation}
which violates the inequality for a wide range of angles. This violation has been confirmed experimentally. Similar inequalities that violate quantum mechanics have been derived for the more general class of hidden variables with
  \begin{equation}
    p^{{\bf a},{\bf b}}_{\mu\nu}(\alpha,\beta)=\frac{1}{4}\Big(1
    								 +\alpha\eta\langle{\bf a}{\bf\sigma}\rangle_\mu
                                                                            +\beta\eta\langle{\bf b}{\bf\sigma}\rangle_\nu
                                                                            +\alpha\beta C_{\mu\nu}({\bf a},{\bf b})\Big)\ ,
     \label{final13}                                                                       
  \end{equation}
where $0<\eta\leq1$. These theories allow for arbitrary correlations and (collectively) for arbitrary non-flat marginals.\footnote{The hidden variables correspond to pairs of density matrices $\eta\mu+(1-\eta)\iden$ and $\eta\nu+(1-\eta)\iden$.}

Thus, Branciard {\em et al}.'s versions of Leggett's theorem are formulated in $2\times 2$ dimensions and show that hidden variables models making non-trivial predictions of a certain kind (pure marginals or pure to degree $\eta$) for the singlet or any sufficiently entangled state and satisfying both PI and CPI  will be incompatible with the predictions of quantum mechanics. The incompatibility is shown by means of Leggett's inequalities, which lead to direct experimental tests. Our theorem is formulated in $3\times 3$ dimensions and shows that any hidden variables model making non-trivial predictions for the singlet state (including models with flat marginals) and satisfying both PI and CPI  will be incompatible with the predictions of quantum mechanics. The incompatibility does not rely on inequalities. The analogy is like that between Bell's original proof of non-locality and those without inequalities by Stairs and others.

If one compares the two theorems (apart from the difference in the dimension of the Hilbert space), the main novelty in our theorem is that it imposes no restriction at all on the form of the marginals. By the same token, however, since the non-triviality assumption is no longer quantified, we lose the possibility of a direct quantitative discrimination between quantum mechanics and hidden variables theories. 

While, as in Leggett's case, our theorem yields new constraints on hidden variables theories that violate the assumptions of Bell's theorem, we believe it also provides a simple explicit classification of which hidden variables theories are possible if one is prepared to violate the relevant assumptions. Any hidden variables theory will either have to violate parameter independence (such as de~Broglie-Bohm theory), or trivialise for the case of maximally entangled states (such as Beltrametti and Bugajski's (1995) canonical classical extension of quantum mechanics, which trivialises for all pure states), or, finally, violate CPI. A hidden variables theory violating this last assumption would exhibit a new and peculiar form of contextuality.

\section*{Appendix: proof of Proposition~4}
We begin by stating a very trivial lemma:

  \noindent{\bf Lemma 1}: {\em For any pure quantum state $\ket{\Psi}$, consider the corresponding mapping 
    $\Phi_{\ket{\Psi}\bra{\Psi}}$ as in Proposition~3. Then for any 1-dimensional projection $Q$ with 
    non-zero probability $p_{\ket{\Psi}}(Q):=\Tr(\ket{\Psi}\bra{\Psi}\iden\otimes Q)$, the positive
    operator $\Phi_{\ket{\Psi}\bra{\Psi}}(Q)$ is also proportional to a 1-dimensional projection.}
    
Indeed, if Bob measures $Q$ and `collapses' the state onto $\iden\otimes Q\ket{\Psi}$, it is well-known 
that the (unnormalised) `collapsed' state on Alice's side is pure.

Next, we prove the lemma that is crucial for the proof: 
 
  \noindent{\bf Lemma 2}: {\em Take any pure quantum state $\ket{\Psi}$, and assume}
      \begin{equation}
        \Phi_{\ket{\Psi}\bra{\Psi}}=\eta\Phi_{\Lambda_1}+(1-\eta)\Phi_{\Lambda_2}\ ,
        \label{finalapp1}
      \end{equation}
  {\em for suitable $\Lambda_k$ ($k=1,2$) defined as in Proposition~3. Then, for any 1-dimensional 
  projection $Q$,}
        \begin{equation}
        \Phi_{\Lambda_k}(Q)=c_k(Q)\Phi_{\ket{\Psi}\bra{\Psi}}(Q)\ ,
        \label{finalapp2}
      \end{equation}
  {\em with proportionality factor $c_k(Q)\geq 0$ that may depend explicitly on $Q$.}
  
Indeed, because of Lemma~1, if $Q$ is a 1-dimensional projection, so (up to a positive multiple) is $\Phi_{\ket{\Psi}\bra{\Psi}}(Q)$. But then it is possible to satisfy
      \begin{equation}
        \Phi_{\ket{\Psi}\bra{\Psi}}(Q)=\eta\Phi_{\Lambda_1}(Q)+(1-\eta)\Phi_{\Lambda_2}(Q)
        \label{finalapp3}
      \end{equation}
only if $\Phi_{\Lambda_1}(Q)$ and $\Phi_{\Lambda_2}(Q)$ are themselves positive multiples of
$\Phi_{\ket{\Psi}\bra{\Psi}}(Q)$, because projection operators generate extremal rays in the convex cone
of positive operators on ${\cal H}$.

With Lemma~2 in hand, we only need to show that $c_k(Q)=1$ for all $Q$ in order to prove Proposition~4.

Now take a basis of 1-dimensional projections $Q_i$ in the space of operators on ${\cal H}$, and
let $Q=\sum_i\alpha_iQ_i$ be an arbitrary 1-dimensional projection. Then, by Lemmas~1 and 2,
  \begin{equation}
    \begin{split}
      \Phi_{\Lambda_k}(Q) & =\Phi_{\Lambda_k}(\sum_i\alpha_iQ_i)=\sum_i\alpha_i\Phi_{\Lambda_k}(Q_i)=\\
                          & =\sum_i\alpha_ic_k(Q_i)\Phi_{\ket{\Psi}\bra{\Psi}}(Q_i)\ .
    \end{split}
  \label{finalapp4}
  \end{equation}

But we similarly know that
  \begin{equation}
    \begin{split}
      \Phi_{\Lambda_k}(Q) & =c_k(Q)\Phi_{\ket{\Psi}\bra{\Psi}}(Q)
                            =c_k(Q)\Phi_{\ket{\Psi}\bra{\Psi}}(\sum_i\alpha_iQ_i)=\\
                          & =c_k(Q)\sum_i\alpha_i\Phi_{\ket{\Psi}\bra{\Psi}}(Q_i)\ .
    \end{split}
  \label{finalapp5}
  \end{equation}
Therefore,
  \begin{equation}
    \sum_i\alpha_ic_k(Q)\Phi_{\ket{\Psi}\bra{\Psi}}(Q_i)=\sum_i\alpha_ic_k(Q_i)\Phi_{\ket{\Psi}\bra{\Psi}}(Q_i)\ .
  \label{finalapp6}
  \end{equation}

We now specialise to the case of a maximally entangled $\ket{\Psi}$, and prove our final
  
  \noindent{\bf Lemma 3}: {\em For any maximally entangled quantum state $\ket{\Psi}$, the mapping 
    $\Phi_{\ket{\Psi}\bra{\Psi}}$, up to a positive factor, is unitary with respect to the Hilbert-Schmidt
    scalar product $\Tr(A^*B)$.}
    
The proof is as follows. Writing a quantum state in biorthogonal form, 
$\ket{\Psi}=\sum_i\beta_i\ket{\psi_i}\ket{\varphi_i}$, we have that
  \begin{equation}
    \begin{split}
      \Tr(\ket{\Psi}\bra{\Psi}P\otimes Q)  &  =\Tr\Big( \Big[\sum_i\beta_i\ket{\psi_i}\ket{\varphi_i}\Big]
                                                                                       \Big[\sum_j\beta_j\bra{\psi_j}\bra{\varphi_j}\Big]
                                                                                       P\otimes Q\Big) =  \\
      &  =\Tr\Big( P\sum_{ij}\beta_i\beta_j\ket{\psi_i}\bra{\psi_j}\bra{\varphi_j}Q\ket{\varphi_i}\Big)\ ,                    
    \end{split}
    \label{final15}
  \end{equation}
so that
  \begin{equation}
    \Phi_{\ket{\Psi}\bra{\Psi}}(Q)=\sum_{ij}\beta_i\beta_j\ket{\psi_i}\bra{\psi_j}\bra{\varphi_j}Q\ket{\varphi_i}\ .
    \label{final16}
  \end{equation}
Now, if $\rho=\ket{\Psi}\bra{\Psi}$ is maximally entangled, on can check explicitly that  $\Phi_{\ket{\Psi}\bra{\Psi}}$ is unitary up to a positive real multiple. Indeed, 
  \begin{multline}
      \Tr\Big(\Phi_{\ket{\Psi}\bra{\Psi}}(A^*)\Phi_{\ket{\Psi}\bra{\Psi}}(B)\Big)  = \\
      \begin{split}
      = & \Tr\Big(\frac{1}{n^2}\sum_{ij}\ket{\psi_j}\bra{\psi_i}\bra{\varphi_i}A^*\ket{\varphi_j}
                     \frac{1}{n^2}\sum_{mn}\ket{\psi_n}\bra{\psi_m}\bra{\varphi_m}B\ket{\varphi_n}\Big)=\\
      = & \frac{1}{n^4}\sum_{ij}\bra{\varphi_i}A^*\ket{\varphi_j}\bra{\varphi_j}B\ket{\varphi_i}
      =  \frac{1}{n^4}\Tr(A^*B)\ .
    \end{split}  
    \label{final17}
  \end{multline}

We can now complete the proof of Proposition~4. Consider again the basis $Q_i$ in the space of operators. 
By Lemma~3, if $\ket{\Psi}$ is maximally entangled, the operators $\Phi_{\ket{\Psi}\bra{\Psi}}(Q_i)$ will 
also form a basis. In this case, therefore, by the uniqueness
of basis expansions in (\ref{finalapp6}) it follows that 
  \begin{equation}
    c_k(Q)=c_k(Q_i)
    \label{finalapp7}
  \end{equation}
for arbitrary 1-dimensional projections $Q$ and all $Q_i$ in the expansion of $Q$. It follows that
  \begin{equation}
    c_k(Q)=c_k\ ,
    \label{finalapp8}
  \end{equation}
independently of $Q$. 
Finally, since $\Tr(\Phi_{\ket{\Psi}\bra{\Psi}}(\iden))=\Tr(\Phi_{\Lambda_k}(\iden))=1$, 
we have $c_k=1$. This completes the proof.\\

{\bf Acknowledgements}\\
The beginning of these investigations goes back to fruitful discussions with Soazig Le~Bihan in 2005--06, and this paper was partly prompted by her work (cf. Le Bihan, 2008). 
More recent discussions with Harvey Brown have been very helpful in clarifying the issues further. For 
the proof of Proposition 4, I am very much indebted to suggestions by and correspondence with Howard 
Barnum and Alex Wilce. I am further very grateful to Matteo Morganti and especially Nicolas Gisin for 
focussing my attention on Leggett's theorem. I wish to thank Stephen Bartlett, Richard Gill, 
Owen Maroney, Huw Price, Rafael Sorkin and Hans Westman for useful discussions and criticism, and, finally, Michiel Seevinck for his very perceptive remarks on the final draft. This paper was written during my tenure of a Senior Research Fellowship at the Centre for Time, Department of Philosophy, University of Sydney, and in part during a research visit at the Perimeter Institute for Theoretical Physics, Waterloo, Ontario. Financial support from the Australian Research Council (Discovery Project 0556184) and the PIAF Collaboration is gratefully acknowledged.\\

{\bf References}\\
Bacciagaluppi, G. (1993), `Critique of Putnam's quantum logic', {\em International Journal of Theoretical Physics} {\bf 32}, 1835--1846.

Barnum, H., Fuchs, C. A.,  Renes, J. M., and Wilce, A. (2005), `Influence-free states on compound quantum systems', arXiv:quant-ph/0507108.

Bell, J. S. (1971), `Introduction to the hidden-variable question', in B. d'Espagnat (ed.), {\em Foundations of Quantum Mechanics}, Proceedings of the International School of Physics `Enrico Fermi', course~IL (New York: Academic), pp.~171--181. Reprinted as Ch.~4 of J.~S.~Bell, {\em Speakable and Unspeakable in Quantum Mechanics} (Cambridge: CUP, 1987), pp.~29-39.

Beltrametti, E., and Bugajski, S. (1995), `A classical extension of quantum mechanics', {\em Journal of Physics} {\bf A 28}, 3329--3344.

Branciard, C., Brunner, N., Gisin, N., Kurtsiefer, C., Lamas-Linares, A., Ling, A., and Scarani, V. (2008), `A simple approach to test Leggett's model of nonlocal quantum correlations', {\em Nature Physics} {\bf 4}, 681--685. Also arXiv:0801.2241. 

Brown, H. R., and Svetlichny, G.  (1990), `Nonlocality and Gleason's lemma. Part I. Deterministic theories', {\em Foundations of  Physics} {\bf 20}, 1379--1387.

Cerf, N.J., Gisin, N., Massar, S., and Popescu, S. (2005), `Simulating maximal quantum entanglement without communication', {\em Physical Review Letters} {\bf 94}, 220403-1--220403-4. Also arXiv:quant-ph/0507120.

Heywood, P., and Redhead, M. L. G. (1983), `Nonlocality and the Kochen-Specker paradox', {\em Foundations of Physics} {\bf 13}, 481--499. 

Jarrett, J. P. (1984), `On the physical significance of the locality conditions in the Bell arguments', {\em No\^{u}s} {\bf XVIII}, 569--589.

Jones, M. R., and Clifton, R. K. (1993), `Against experimental metaphysics', {\em Midwest Studies in Philosophy} {\bf XVIII}, 295--316.

Le Bihan, S. (2008), {\em Understanding Quantum Phenomena}, Ph.D. dissertation, University of Nancy 2 and University of Bielefeld. 

Leggett, A. J. (2003), `Nonlocal hidden-variable theories and quantum mechanics: an incompatibility theorem', {\em Foundations of Physics} {\bf 33}, 1469--1493. 

Shimony, A. (1978), `Metaphysical problems in the foundations of quantum mechanics', {\em International Philosophical Quarterly} {\bf 18}, 3--17.

Shimony, A. (1986), `Events and processes in the quantum world', in R. Penrose and C. Isham (eds.), {\em Quantum Concepts in Space and Time} (Oxford: Clarendon Press), pp.~182--203. Repr. in A. Shimony, {\em Search for a Naturalistic World View}, Vol.~II (Cambridge: CUP, 1993), pp.~140--162.

Stairs, A. (1983), `Quantum logic, realism, and value definiteness', {\em Philosophy of Science} {\bf 50}, 578--602.

\end{document}